\UseRawInputEncoding

\documentclass[11pt]{article}
\usepackage{longtable,supertabular}
\usepackage{amsmath}
\usepackage{amssymb}
\usepackage{latexsym}
\usepackage{cite}
\usepackage{cite,color}
\usepackage{bm}

\title{\bf Construction and Fast Decoding of Binary Linear Sum-Rank-Metric Codes}
\author{Hao Chen, Yanfeng Qi and Zhiqiang Cheng
  \thanks{Hao Chen and Zhiqiang Cheng are with the College of Information Science and Technology/Cyber Security, Jinan University, Guangzhou, Guangdong Province, 510632, China, haochen@jnu.edu.cn, 2712468769@qq.com.  	
Yanfeng Qi is with School of Science, Hangzhou Dianzi University, Hangzhou, 310018, China, qiyanfeng07@163.com.  	
The research of Hao Chen was supported by NSFC under Grant 62032009.
The research of Yanfeng Qi was supported by
the Zhejiang provincial Natural Science Foundation of China (No. LY21A010013) and
Scientific Research Fund of Zhejiang  Provincial Education Department (No.Y202249655).
}}

\begin{document}

\maketitle
\begin{abstract}
Sum-rank-metric codes have wide applications in the multishot network coding and the distributed storage. Linearized Reed-Solomon codes, sum-rank BCH codes and their Welch-Berlekamp type decoding algorithms have been proposed and studied. They are sum-rank versions of Reed-Solomon codes and BCH codes in the Hamming-metric. In this paper, we construct binary linear sum-rank-metric codes with the matrix size $2 \times 2$, from BCH, Goppa and additive quaternary Hamming-metric codes. A reduction of the decoding in the binary sum-rank-metric space to the decoding in the  Hamming-metric space is given. Fast decoding algorithms of binary linear BCH-type and Goppa-type  sum-rank-metric codes with the block length $\ell$ and the matrix size $2 \times 2$, which are better than these sum-rank BCH codes, are presented. These fast decoding algorithms for binary linear BCH-type and Goppa-type sum-rank-metric codes with the matrix size $2 \times 2$ need at most $O(\ell^2)$ operations in the field ${\bf F}_4$. Asymptotically good sequences of quadratic-time encodable and decodable binary linear sum-rank-metric codes with the matrix size $2 \times 2$ can be constructed from Goppa codes.\\

\end{abstract}

\section{Introduction}

{\color{red}\subsection{Preliminaries}}

Sum-rank-metric codes have wide applications in multishot network coding, see \cite{MK19,NPS,NU}, space-time coding, see \cite{SK}, and coding for distributed storage,  see \cite{CMST,MK,MP1}. There have been many papers on the structures, constructions and applications of sum-rank-metric codes in recent years, see e.g.\cite{BGR1,BGR,MP1,MP191,MK19,MP20,MP21,Neri,NSZ21,MP22} and references therein. Fundamental properties and some bounds on sizes of sum-rank-metric codes were given in the paper \cite{BGR}. For a nice survey of sum-rank-metric codes and their applications, we refer to \cite{MPK22}.\\

We recall some basic concepts and results for sum-rank-metric codes in \cite{BGR}. Let ${\bf F}_q^{n\times m}$ be the set of all $ n \times m$ matrices. This is a linear space over ${\bf F}_q$ of the dimension $nm$. Let $n_i \leq m_i$ be $2\ell$ positive integers satisfying $m_1 \geq m_2 \cdots \geq m_\ell$. Set $N=n_1+\cdots+n_\ell$.  Let $${\bf F}_q^{n_1 \times m_1, \ldots,n_\ell\times m_\ell}={\bf F}_q^{n_1 \times m_1} \bigoplus \cdots \bigoplus {\bf F}_q^{n_\ell \times m_\ell}$$ be the set of all ${\bf x}=({\bf x}_1,\ldots,{\bf x}_\ell)$, ${\bf x}_i \in {\bf F}_q^{n_i \times m_i}$, $i=1,\ldots,\ell$. This is a linear space over ${\bf F}_q$ of the dimension $\Sigma_{i=1}^\ell n_im_i$. {\color{red}$n_1 \times m_1, \ldots, n_\ell \times m_\ell$ are called matrix sizes of the sum-rank-metric code. For $i=1,2,..., \ell$, if each $n_i$ is equal and each $m_i$ is equal, then the matrix size of this code is simply $n_1\times m_1$.} Set $wt_{sr}({\bf x}_1, \ldots, {\bf x}_\ell)=rank({\bf x}_1)+\cdots+rank({\bf x}_\ell)$ and $$d_{sr}({\bf x},{\bf y})=wt_{sr}({\bf x}-{\bf y}),$$ for ${\bf x}, {\bf y} \in {\bf F}_q^{n_1\times m_1, \ldots,n_\ell\times m_\ell}$. This is indeed a metric on ${\bf F}_q^{n_1\times m_1, \ldots,n_\ell\times m_\ell}$. \\

{\bf Definition 1.1.} {\em A $q$-ary linear  sum-rank-metric code ${\bf C}$ with block length $\ell$ and matrix sizes $n_1 \times m_1, \ldots, n_\ell\times m_\ell$ is a subset of  the finite matrix space ${\bf F}_q^{n_1\times m_1, \ldots,n_\ell\times m_\ell}$. Its minimum sum-rank distance is defined by $$d_{sr}({\bf C})=\min_{{\bf x} \neq {\bf y}, {\bf x}, {\bf y} \in {\bf C}} d_{sr}({\bf x}, {\bf y}).$$  The code rate of ${\bf C}$ is $R_{sr}=\frac{log_q |{\bf C}|}{\Sigma_{i=1}^\ell n_im_i}$. The relative distance is $\delta_{sr}=\frac{d_{sr}({\bf C})}{N}$. Without causing confusion, $d_{sr}({\bf C})$ is abbreviated as $d_{sr}$ in the rest of this paper. When ${\bf C} \subset {\bf F}_q^{n_1\times m_1, \ldots,n_\ell\times m_\ell}$ is a linear subspace of the linear space ${\bf F}_q^{n_1\times m_1, \ldots,n_\ell\times m_\ell}$ over ${\bf F}_q$, ${\bf C}$ is called a linear sum-rank-metric code. When $q=2$, this code is called binary sum-rank-metric code.}\\

In general it is interesting to construct good sum-rank-metric codes with large cardinalities and large minimum sum-rank distances, and develop their decoding algorithms. For $\ell=1$, this is the rank-metric code case, see \cite{Gabidulin}. For $m_1=\cdots=m_\ell=m$, $n_1=\cdots=n_\ell=n$, this is the $m$-sum-rank-metric code over ${\bf F}_{q^m}$ with the length $n\ell$. For $m=n=1$, this is the Hamming-metric code case. Hence the sum-rank-metric is a generalization and combination of the Hamming-metric and the rank-metric. When $q=2$, $n_1=\cdots=n_\ell=m_1=\cdots=m_\ell=2$, sum-rank-metric codes are in the space ${\bf F}_2^{2\times 2, \ldots, 2\times 2}={\bf F}_2^{2\times 2} \bigoplus \cdots \bigoplus {\bf F}_2^{2 \times 2}$, this is the case combining binary Hamming-metric codes and $2 \times 2$ binary matrices. In this paper we concentrate on the construction and the fast decoding of binary linear sum-rank-metric codes in ${\bf F}_2^{2\times 2, \ldots, 2\times 2}$.\\

When the matrix size $n \times m$ is fixed, a sequence of sum-rank-metric codes ${\bf C}_i$ of the block size $\ell_i \longrightarrow \infty$, satisfying $$R_{sr}=\lim \frac{log_q |{\bf C}_i|}{mn\ell_i}>0$$ and $$\delta_{sr}=\lim\frac{d_{sr}({\bf C}_i)}{n\ell_i}>0,$$ is called an asymptotically good sequence of sum-rank-metric codes. It is interesting to get asymptotic {\color{red}bounds on $R_{sr}$ and $\delta_{sr}$.}\\

The Hamming weight $wt_H({\bf a})$ of a vector ${\bf a}=(a_1, \ldots, a_n) \in {\bf F}_q^n$ is the number of coordinate positions in its support $$supp({\bf a})=\{i:a_i \neq 0\}.$$ The Hamming distance $d_H({\bf a}, {\bf b})$ between two vectors ${\bf a}$ and ${\bf b}$ is defined as the Hamming weight $wt_H({\bf a}-{\bf b})$. For a code ${\bf C} \subset {\bf F}_q^n$ of dimension $k$, its Hamming distance is $$d_H=\min_{{\bf a} \neq {\bf b}} \{d_H({\bf a}, {\bf b}): {\bf a}, {\bf b} \in {\bf C}\}.$$ It is well-known that the Hamming distance of a linear code ${\bf C}$ is the minimum Hamming weight of its non-zero codewords. For the theory of error-correcting codes in the Hamming-metric, we refer to the books \cite{MScode,Lint,HP}.   A code ${\bf C}$ in ${\bf F}_q^n$ of the minimum distance $d_H$ and the cardinality $M$ is denoted by an $(n, M, d_H)_q$ code. The code rate is $$R({\bf C})=\frac{log_q M}{n},$$ and the relative distance is $$\delta({\bf C})=\frac{d_H}{n}.$$ For a linear $[n, k, d_H]_q$ code, the Singleton bound asserts $d_H \leq n-k+1$. When equality holds, this code is called a maximal distance separable (MDS) code. And Reed-Solomon codes are well-known MDS codes, see e.g.\cite{HP}. BCH codes and Goppa codes can be considered as subfield subcodes of generalized Reed-Solomon codes, and are severed as optimal binary codes for many parameters, see \cite{MScode,HP,codetable}.\\

Each Hamming-metric $[\ell, k, d_H]_{q^n}$ code can be considered as a sum-rank-metric code of the matrix size $n \times n$ and  the block length $\ell$ over ${\bf F}_q$, by taking each coordinate in ${\bf F}_{q^n}$ as the first row of the $n \times n$ matrix, and padding zero other rows to this $n \times n$ matrix. This is a dimension $nk$ sum-rank-metric code, with the minimum sum-rank distance at least $d_H$. Then the code rate of this sum-rank-metric code is $$R_{sr}=\frac{nk}{n^2\ell}=\frac{R({\bf C})}{n}$$ and $$\delta_{sr}=\frac{d_H}{n\ell}=\frac{\delta({\bf C})}{n}.$$ Each Hamming-metric $[n\ell, k, d_H]_{q^n}$ code can be considered as a sum-rank-metric code with the matrix size $n \times n$ and  the block length $\ell$ over ${\bf F}_q$. By taking every consecutive $n$ coordinates in ${\bf F}_{q^n}$ as the $n \times n$ matrix. This is a dimension $nk$ sum-rank-metric code, with the minimum sum-rank distance at least $\frac{d_H}{n}$. Then the code rate of this sum-rank-metric code is $$R_{sr}=\frac{nk}{n^2\ell}=R({\bf C})$$ and $$\delta_{sr} \geq \frac{d_H}{n^2\ell}=\frac{\delta({\bf C})}{n}.$$\\

Therefore asymptotically good sequences of sum-rank-metric codes of the matrix size $n \times n$ can be obtained directly from asymptotically good sequence of Hamming-metric codes meeting the Gilbert-Varshamov bound. The code rate and the relative distance of this sequence of sum-rank-metric codes satisfy $$ R_{sr} \geq 1-H_{q^n}(n\delta_{sr}),$$ where $H_{q^n}(x)$ is the $q^n$-ary entropy function. The $q$-ary entropy function is defined on $0\leq x \leq 1-\frac{1}{q}$ by $$H_q(0)=0$$ and $$H_q(x)=xlog_q(q-1)-xlog_qx-(1-x)log_q(1-x),$$ for $x \neq 0$. Asymptotically good sequences of quadratic-time encodable and decodable sum-rank-metric codes obtained in Section 4 have much better asymptotic parameters than the above bound.\\

The rank-metric on the space ${\bf F}_q^{n \times m}$ of size $n \times m$ matrices over ${\bf F}_q$, where $n \leq m$, is defined by the ranks of matrices, $d_r(A,B)= rank(A-B)$. The minimum rank-distance of a code ${\bf C} \subset {\bf F}_q^{n \times m}$  is $$d_r({\bf C})=\min_{A\neq B} \{d_r(A,B): A, B \in {\bf C} \}.$$  For a code ${\bf C}$ in ${\bf F}_q^{n \times m}$ with the minimum rank distance $d_r({\bf C}) = d$, the Singleton-like bound asserts that the number of codewords in ${\bf C}$ is upper bounded by $q^{m(n-d+1)}$ , see \cite{Gabidulin}.  A code satisfying the equality is called a maximal rank distance (MRD) code. The Gabidulin code in ${\bf F}_q^{n\times n}$ is consisting of ${\bf F}_q$-linear mappings on ${\bf F}_q^n \backsimeq {\bf F}_{q^n}$ defined by $q$-polynomials $a_0x+a_1x^q+\cdots+a_ix^{q^i}+\cdots+a_\ell x^{q^\ell}$, where $a_\ell,\ldots,a_0$ are arbitrary elements in ${\bf F}_{q^n}$, see \cite{Gabidulin}. Then the rank-distance of the Gabidulin code is at least $n-\ell$ since each above $q$-polynomial has at most $q^\ell$ roots in ${\bf F}_{q^n}$. There are $q^{n(\ell+1)}$ such $q$-polynomials. Hence the size of the Gabidulin code is $q^{n(\ell+1)}$ and it is an MRD code. \\

The Singleton-like bound for the sum-rank-metric was proposed in \cite{MK,BGR}. The general form Theorem III.2 in \cite{BGR} is as follows. The minimum sum-rank distance $d$ can be written uniquely as the form $d_{sr}=\Sigma_{i=1}^{j-1} n_i+\delta+1$ where $0 \leq \delta \leq n_j-1$, then $$|{\bf C}| \leq q^{\Sigma_{i=j}^\ell n_im_i-m_j\delta}.$$ The code attaining this bound is called a maximal sum-rank-metric distance (MSRD) code. When $m_1=\cdots=m_\ell=m,$ this bound is of the form $$|{\bf C}| \leq q^{m(N-d_{sr}+1)}.$$

\subsection{Related works and our contributions}

Constructions of MSRD codes attaining the Singleton-like bound are important and interesting, we refer to \cite{BGR,MP20,MPS24,MP24,Chen}. The first infinite family of MSRD codes are linearized Reed-Solomon codes introduced and studied in \cite{MP1}, which are the sum-rank versions of the classical Reed-Solomon codes. Their Welch-Berlekamp decoding algorithm was presented in \cite{MK19}. Twisted version of linear Reed-Solomon codes was give in \cite{Neri}. In \cite{BGR}, some MSRD codes were constructed and their constructions were generalized in \cite{Lao1}. In \cite{MP20,Chen}, MSRD codes with various matrix sizes were constructed. MSRD codes over finite chain rings were constructed in a recent paper \cite{MPS24}. New constructions and generalizations of previous constructions of MSRD codes were given in the recent paper \cite{MP24}. It was proved in \cite{BGR,AKR,Chen2} that there is no MSRD code for some parameters.\\

There are some constructions of sum-rank-metric codes with good parameters over small fields, we refer to the subfield subcodes of linearized Reed-Solomon codes in \cite{MP1} and sum-rank BCH codes in \cite{MP21}. Sum-rank BCH codes of the matrix size $n_1=\cdots=n_\ell=n$, $m_1=\cdots=m_\ell=m$ were proposed and studied in \cite{MP21} by the deep algebraic method. These sum-rank-metric codes are linear over ${\bf F}_{q^m}$.  There is a designed distance of such a sum-rank BCH code such that the minimum sum-rank distance is greater than or equal to the designed distance.  On the other hand the dimensions of these sum-rank BCH codes are lower bounded in \cite{MP21} Theorem 9. Many sum-rank-metric codes of the parameters $n=m=2$ and $q=2$ were constructed in Tables I, II, III, IV, V, VI and VII of \cite{MP21}. The minimum sum-rank distances of these sum-rank BCH codes were improved in a recent paper \cite{ALNWZ}. The decoding algorithm for sum-rank BCH codes was presented in \cite{MP21}. In our previous paper \cite{Chen}, we proposed a new construction of linear sum-rank-metric codes by combining several Hamming-metric codes and $q$-polynomial representations of rank-metric codes. Most of constructed sum-rank-metric codes in the paper \cite{Chen} from presently known best linear codes in \cite{codetable} are larger than the sum-rank BCH codes in \cite{MP21} of the same minimum {\color{red}sum-rank} distances. No decoding algorithm has been developed for sum-rank-metric codes in \cite{Chen}.\\

Fast algebraic decoding and list-decoding algorithms for BCH codes, Goppa codes and algebraic geometry codes in the Hamming-metric are well-developed, see \cite[Chapter 5]{HP}, \cite{Feng,Chen1} and \cite{GS}. It is well-known that BCH codes and Goppa codes over ${\bf F}_q$ in the Hamming-metric can be decoded up to $\left \lfloor \frac{d-1}{2}\right \rfloor$ errors within the cost of $O(\ell^2)$ operations of the field ${\bf F}_q$, where $\ell$ is the length of the code, we refer to \cite[Chapter 6]{Lint}, \cite[Chapter 5]{HP} and \cite[Chapter 9, Chapter 12]{MScode}. To the best of our knowledge, there are few known decoding algorithms of sum-rank-metric codes. The fundamental decoding algorithms in the sum-rank-metric are the Welch-Berlekamp decoding algorithm for linearized Reed-Solomon codes given in \cite{MK19} and the decoding algorithm for the sum-rank BCH codes given in \cite{MP21}. The complexity of their decoding algorithms is $O(\ell^2)$ operations in the field ${\bf F}_{q^m}$. For the fast decodings of linearized Reed-Solomon codes, interleaved linearized Reed-Solomon codes and folded linearized Reed-Solomon codes, we refer to \cite{Bartz,Bartz2,Bartz3,Bartz4}. Generic decoder in the sum-rank-metric was developed in \cite{Puchinger}. \\

Linearized Reed-Solomon codes were extended to algebraic geometry sum-rank codes in \cite{Berardini}, Codes with few nonzero sum-rank weights and minimal sum-rank codes have been studied in \cite{NSZ21,Borello}.\\

Additive quaternary codes were studied from the motivation of finite geometries and the construction of quantum codes in \cite{Blokhuis,Bier,Bier1,Bier2}. Additive quaternary codes are generalizations of quaternary linear codes.\\

The first contribution of this paper is as follows. We construct binary linear sum-rank-metric codes with the matrix size $2 \times 2$ from quaternary BCH and Goppa codes. These binary linear sum-rank-metric codes are called BCH and Goppa-type sum-rank-metric codes. Many BCH and Goppa-type binary linear sum-rank-metric codes are larger than the sum-rank BCH codes constructed by U. Mart\'{\i}nez-Pe\~{n}as in \cite{MP1,MP21} with the same minimum sum-rank distances. We construct binary sum-rank-metric codes with matrix size $2\times 2$ from two additive quaternary codes. They are generalizations of binary sum-rank-metric codes with matrix size $2\times 2$ constructed from two quaternary linear codes. Furthermore, binary sum-rank-metric codes with matrix size $2\times 2$ constructed from two additive quaternary codes are still linear. Thus we show by examples that binary linear sum-rank-metric codes with matrix size $2\times 2$ constructed from two additive quaternary codes are superior to that constructed from two of the best known quaternary linear codes. These constructions are mainly based on the construction of \cite{Chen}.\\

More importantly, we {\color{red}give a reduction} of the decoding in the binary sum-rank-metric space to the decoding in the Hamming-metric space. Then this reduction is applied to BCH and Goppa-type binary linear sum-rank-metric codes constructed quaternary Hamming-metric BCH and Goppa codes. The reduction is based on the following main result 1.  Then from various fast decoding algorithms of quaternary BCH and Goppa codes, we present fast decoding algorithms of our BCH and Goppa-type binary linear sum-rank-metric codes. The complexity is $O(\ell^2)$ operations in the field ${\bf F}_{q^2}={\bf F}_4$. Hence our decoding algorithm has the same complexity as the decoding algorithm presented in \cite{MP21}. However our BCH-type sum-rank-metric codes are larger than these sum-rank BCH codes with the same minimum sum-rank distances. Our reduction of decoding is general and then can be applied to binary linear sum-rank-metric codes with the matrix size $2 \times 2$ constructed from any quaternary Hamming-metric codes with the fast decoding.\\

The third contribution is the construction of asymptotically good sequences of binary linear sum-rank-metric codes with the matrix size $2 \times 2$ from Goppa codes, which can be decoded within the $O(\ell^2)$ operations of the field ${\bf F}_4$.\\

	For the convenience of writing, set
	$${\bf a}_nx+...+{\bf a}_{1}x^{q^{n-1}}=\left(a_{1,n}x+...+a_{1,1}x^{q^{n-1}},...,a_{\ell,n}x+...+a_{\ell,1}x^{q^{n-1}}\right)$$ $\in {\bf F}_q^{\overbrace {n\times n,...,n\times n}^{\ell}},$
	where ${\bf a}_i=(a_{1,i}, \ldots, a_{\ell,i})\in {\bf F}_{q^n}^{\ell}, i=1,...,n.$ The linear sum-rank-metric code $SR({\bf C}_1, {\bf C}_2)$ of the matrix size $2 \times 2$ constructed from two quaternary codes ${\bf C}_i \subset {\bf F}_4^{\ell}$, $i=1,2 $, is defined by $$SR({\bf C}_1,{\bf C}_2)=\{{\bf a}_2x+{\bf a}_{1}x^2: {\bf a}_1=(a_{1,1},\ldots, a_{\ell,1}) \in {\bf C}_1, {\bf a}_2=(a_{1,2},\ldots, a_{\ell,2}) \in {\bf C}_2\}.$$

	{\bf Main result 1.} {\em Let ${\bf C}_1 \subset {\bf F}_4^\ell$ and ${\bf C}_2 \subset {\bf F}_4^\ell$ be two linear $[\ell, k_1, d_1]_4$ and $[\ell, k_2, d_2]_4$ codes over ${\bf F}_4$. Then a binary linear sum-rank-metric code $SR({\bf C}_1, {\bf C}_2)$ with block length $\ell$ and matrix size $2 \times 2$ can be constructed explicitly. The dimension of $SR({\bf C}_1, {\bf C}_2)$ over ${\bf F}_2$ is $\dim_{{\bf F}_2}(SR({\bf C}_1, {\bf C}_2))=2(k_1+k_2)$. For any two codewords, ${\bf a}_1 \in {\bf C}_1$ and ${\bf a}_2 \in {\bf C}_2$, set $$I=supp({\bf a}_1) \cap supp({\bf a}_2),$$ then $$wt_{sr}({\bf a}_2x+{\bf a}_1x^2)=2wt_H({\bf a}_1)+2wt_H({\bf a}_2)-3|I|.$$ It follows that $d_{sr}(SR({\bf C}_1, {\bf C}_2))=d_{sr}(SR({\bf C}_2, {\bf C}_1))$ is at least $$\max \{\min\{d_1, 2d_2\},\min\{d_2, 2d_1\}\}.$$}
	
	In the main result 1 above,  the sum-rank weight of each codeword in binary linear sum-rank-metric codes with the matrix size $2\times 2$ is computed exactly from Hamming weights of two quaternary linear Hamming-metric codes. Compared to Construction 2 of our previous paper \cite{Chen}, the lower bound on the minimum sum-rank distance is improved. This result provides a technique for the fast decoding of binary linear sum-rank-metric codes with the matrix size $2\times 2$ in this paper.\\

From the reduction of decodings based on the main result 1, we have the following two results.\\

{\bf Main result 2.} {\em By the usage of two quaternary BCH codes, explicit block length $4^h-1$, $h=2, 3, \ldots$, and matrix size $2 \times 2$ binary linear sum-rank-metric codes, which are larger than the sum-rank BCH codes with the same minimum sum-rank distances, can be constructed. These BCH-type binary linear sum-rank-metric codes can be decoded up to $\left \lfloor \frac{d_{sr}-1}{2} \right \rfloor$ errors, within the complexity of $O(\ell^2)$ operations in the field ${\bf F}_4$.}\\

{\bf Main result 3.} {\em By the usage of two Goppa codes, asymptotically good sequence of binary linear sum-rank-metric codes with the matrix size $2 \times 2$ satisfying $$R_{sr}(\delta_{sr}) \geq 1-\frac{1}{2}(H_4(\frac{4}{3}\delta_{sr})+H_4(2\delta_{sr})),$$ which can be efficiently encoded and decoded within $O(\ell^2)$ operations of the field ${\bf F}_4$, are constructed.}\\

The class of linear sum-rank-metric codes with the matrix size $2 \times 2$ is the first natural extension of linear error-correcting codes in the Hamming-metric, which can be considered as linear sum-rank-metric codes with the matrix size $1 \times 1$. In this paper we concentrate on the construction and decoding of linear sum-rank-metric codes with the matrix size $2 \times 2$. In the matrix size $1 \times 1$ case, the main results in \cite{Spielman} gave the construction of asymptotically good sequence of linear-time encodable and decodable error-correcting codes. \\

\section{BCH-type and Goppa-type sum-rank-metric codes}

In our previous paper \cite{Chen}, the linear space ${\bf F}_q^{n\times n}$ of all $n \times n$ matrices over ${\bf F}_q$ can be identified with the space of all $q$-polynomials, $${\bf F}_q^{n\times n}\backsimeq \{a_0x+a_1x^q+\cdots+a_{n-1}x^{q^{n-1}}: a_0, \ldots, a_{n-1} \in {\bf F}_{q^n}\}.$$ This is the isomorphism of the linear space over ${\bf F}_q$.\\

The construction of $q$-ary linear sum-rank-metric codes with the matrix size $n \times n$ from $n$ linear $q^n$-ary codes is as follows. The first linear $[\ell, k_1, d_1]_{q^{n}}$ code ${\bf C}_1 \subset {\bf F}_{q^{n}}^\ell$ corresponds to the coefficients of $x^{q^{n-1}}$ in the $q$-polynomial. The second linear $[\ell, k_2, d_2]_{q^{n}}$ code ${\bf C}_2 \subset {\bf F}_{q^{n}}^\ell$ corresponds to the coefficients of $x^{q^{n-2}}$ in the $q$-polynomial and so on. Then we can obtain a subset $$SR({\bf C}_1,...,{\bf C}_n)=\{{\bf a}_nx+...+{\bf a}_{1}x^{q^{n-1}}: {\bf a}_n=(a_{1,n},\ldots, a_{\ell,n}) \in {\bf C}_n,...,$$ ${\bf a}_1=(a_{1,1}, \ldots, a_{\ell,1}) \in {\bf C}_1\}$ of ${\bf F}_q^{\overbrace {n\times n,...,n\times n}^{\ell}}.$ This is a $q$-ary linear sum-rank-metric code with matrix size $n\times n$ and  minimum sum-rank distance at least $\min \{d_1, 2d_2,...,nd_n\}$. It is easy to verify the linear independence, therefore the dimension $\dim_{{\bf F}_q}(SR({\bf C}_1,...,{\bf C}_n))=n(k_1+...+k_n)$. The following construction is a special case of our earlier paper \cite{Chen} Construction 2 with $q = 2, v = 1$ and matrix size $2\times 2.$ However, we provide a more detailed proof in this paper.\\

{\bf Theorem 2.1.} {\em (\cite[Theorem 2.1]{Chen}). Let ${\bf C}_1 \subset {\bf F}_4^\ell$ and ${\bf C}_2 \subset {\bf F}_4^\ell$ be two linear $[\ell, k_1, d_1]_4$ and $[\ell, k_2, d_2]_4$ codes over ${\bf F}_4$. Then a binary linear sum-rank-metric code $SR({\bf C}_1, {\bf C}_2)$ with block length $\ell$, matrix size $2 \times 2$ and minimum sum-rank distance at least $\min\{d_1, 2d_2\}$ can be explicitly constructed. The dimension of $SR({\bf C}_1, {\bf C}_2)$ over ${\bf F}_2$ is $\dim_{{\bf F}_2}(SR({\bf C}_1, {\bf C}_2))=2(k_1+k_2)$.}\\

{\bf Proof.} It is clear that the dimension of the linear sum-rank-metric code $SR({\bf C}_1, {\bf C}_2)$ over ${\bf F}_2$ is $2(k_1+k_2)$. For any nonzero codeword ${\bf c}={\bf a}_2x+{\bf a}_1x^2 \in SR({\bf C}_1, {\bf C}_2)$, if ${\bf a}_1 \in {\bf C}_1$ is nonzero, there are at least $d_1$ positions such that the $2 \times 2$ coordinate matrix is not zero. Then $wt_{sr}({\bf c})\geq d_1$. Otherwise if ${\bf a}_1$ is a zero codeword, then ${\bf c}={\bf a}_2 x$, it follows that $wt_{sr}({\bf c})=2wt_H({\bf a}_2) \geq 2d_2$. The conclusion is proved.\\

Below we state for the first time that the sum-rank weight of each codeword in $SR({\bf C}_1, {\bf C}_2)$ can be computed exactly from the corresponding two Hamming-metric codes. This will play a crucial role in the decoding of the binary linear sum-rank-metric codes with matrix size $2\times 2$ in Section 4.\\

{\bf Theorem 2.2.} {\em Let ${\bf C}_1 \subset {\bf F}_4^\ell$ and ${\bf C}_2 \subset {\bf F}_4^\ell$ be two linear $[\ell, k_1, d_1]_4$ and $[\ell, k_2, d_2]_4$ codes over ${\bf F}_4$. Then a binary linear sum-rank-metric code $SR({\bf C}_1, {\bf C}_2)$ with block length $\ell$ and matrix size $2 \times 2$ can be constructed explicitly. For any two codewords, ${\bf a}_1 \in {\bf C}_1$ and ${\bf a}_2 \in {\bf C}_2$, set $$I=supp({\bf a}_1) \cap supp({\bf a}_2),$$ then $$wt_{sr}({\bf a}_2x+{\bf a}_1x^2)=2wt_H({\bf a}_1)+2wt_H({\bf a}_2)-3|I|.$$}

{\bf Proof.} For the coordinate position $i \in supp({\bf a}_1)\backslash supp({\bf a}_2)$, it is clear the matrix ${\bf a}_2x+{\bf a}_1x^2$ in this position is of the rank $2$. For the coordinate position $i \in supp({\bf a}_2)\backslash supp({\bf a}_1)$, it is clear the matrix ${\bf a}_2x+{\bf a}_1x^2$ in this position is of the rank $2$. For these coordinate positions $i \in I$, it is clear that the matrix is of rank $1$, since there is one root $x=\frac{a_{i,1}}{a_{i,2}}$, where $${\bf a}_1=(a_{1,1}, \ldots, a_{\ell,1}),$$ $${\bf a}_2=(a_{1,2}, \ldots, a_{\ell,2}).$$ Then it follows that $$wt_{sr}({\bf a}_2x+{\bf a}_1x^2)=2(wt_H({\bf a}_1)-|I|)+2(wt_H({\bf a}_1)-|I|)+|I|.$$ The conclusion follows immediately.\\

{\bf Remark 2.1.} From Theorem 2.2, if the coefficients of $x$ and $x^2$ are swapped, their sum-rank weights remain the same. It can be concluded that the minimum sum-rank distances of $SR({\bf C}_1, {\bf C}_2)$ and $SR({\bf C}_2, {\bf C}_1)$ are the same. It follows that the minimum sum-rank distances of $SR({\bf C}_1, {\bf C}_2)$ and $SR({\bf C}_2, {\bf C}_1)$ are at least $$\max \{\min\{d_1, 2d_2\},\min\{d_2, 2d_1\}\}.$$ The following example show that the two resulting binary linear sum-rank-metric codes are generally not equivalent.\\

{\bf Example 2.1.}	 Let ${\bf C}_1$ be the linear $[3, 1, 3]_4$ code over ${\bf F}_4$ with generator matrix $[1, \omega, \omega^2],$ where $\omega$ is the primitive element of the field ${\bf F}_4$. Let ${\bf C}_2$ be the linear $[3, 0, 0]_4$ code over ${\bf F}_4.$ Assume $SR({\bf C}_1, {\bf C}_2)$ is equivalent to $SR({\bf C}_2, {\bf C}_1)$. Then there exists a bijection $\alpha_1x+\alpha_2x^2$ between $SR({\bf C}_1, {\bf C}_2)$ and $SR({\bf C}_2, {\bf C}_1)$, where $\alpha_1,\alpha_2\in {\bf F}_4$. It follows from ${\bf C}_2\equiv {\bf 0}$ that $\alpha_1 = 0.$ Hence we have the mapping $\alpha_2x^2(\alpha_2\not=0): SR({\bf C}_1, {\bf C}_2)\rightarrow SR({\bf C}_2, {\bf C}_1)$ by ${\bf a}_1x^2 \longmapsto \alpha_2{\bf a}_1^2x,$ a contradiction since $\alpha_2{\bf a}_1^2 \not\in {\bf C}_1$ for arbitrary ${\bf a}_1\in {\bf C}_1\backslash\{{\bf 0}\}.$ It follows that $SR({\bf C}_1, {\bf C}_2)$ and $SR({\bf C}_2, {\bf C}_1)$ are not equivalent.

\subsection{BCH-type sum-rank-metric codes}

When both ${\bf C}_1$ and ${\bf C}_2$ are BCH codes, we call the constructed binary linear sum-rank-metric codes BCH-type sum-rank-metric codes, for distinguishing from these sum-rank BCH codes in \cite{MP21}. Actually, when both ${\bf C}_1$ and ${\bf C}_2$ are cyclic code, then for each codeword $(A_1, \ldots, A_\ell) \in {\bf F}_2^{2\times 2, \ldots, 2\times 2}$ in the constructed sum-rank-metric code in Theorem 2.1, where $A_i$, $i=1, 2, \ldots, \ell$, is a $2 \times 2$ matrix over ${\bf F}_2$, $(A_\ell, A_1, \ldots, A_{\ell-1})$ is clearly another codeword. By the comparison with Definition 4 in \cite{MP21}, this satisfies one requirement of cyclic-skew-cyclic (CSC) sum-rank-metric code in \cite{MP21}. It seems that the requirement of CSC sum-rank-metric codes are stronger and this is the reason that our sum-rank-metric codes are larger than these CSC codes with same minimum sum-rank distances in \cite{MP21}.\\

Notice that there are several possibilities to take binary or quaternary BCH codes in Theorem 2.1. We can apply Theorem 2.1 to two binary BCH codes, two quaternary BCH codes, or one binary BCH code and one quaternary BCH code. It is interesting to observe that the resulted linear sum-rank-metric codes from quaternary BCH codes are much larger than sum-rank BCH codes of the same minimum sum-rank distances invented by Mart\'{\i}nez-Pe\~{n}as in \cite{MP21}. Sometimes even our constructions in  Theorems  2.1 and 2.2 are applied to two binary BCH codes, the resulted linear sum-rank-metric codes are larger than sum-rank BCH codes of the same minimum sum-rank distances.\\

For any $s \geq 0$, let $C_s = \{s, sq, sq^2
, ... , sq^{m_s-1}\}$ denote the $q$-cyclotomic coset
containing $s$ modulo $\ell$, where $m_s$ is the least positive integer such that $sq^{m_s} \equiv s({\rm mod} ~\ell)$.\\

{\bf Example 2.2.} We can use two BCH codes over ${\bf F}_4$. Let ${\bf C}_2$ be the length $63$ primitive narrow-sense BCH code over ${\bf F}_4$ with the defining set ${\bf T}_2=C_0\cup C_1\cup C_2\cup C_3\cup C_5$. This is a linear $[63,50,7]_4$ code ${\bf C}_2$. Another BCH code ${\bf C}_1$ over ${\bf F}_4$ with the defining set ${\bf T}_1=C_0\cup C_1\cup C_2\cup C_3\cup C_5\cup C_6\cup C_7\cup C_9\cup C_{10}\cup C_{11}$. This is a linear $[63,35,14]_4$ code. From Theorem 2.1, a binary linear sum-rank-metric code with block length $63$, matrix size $2 \times 2$ and the minimum sum-rank distance at least $14$ is constructed. The dimension of this code is $2 \cdot (50+35)=2 \cdot 85$. This is much larger than the dimension $2 \cdot 70$ sum-rank BCH code of the same minimum sum-rank distance $14$ in \cite[Table VI]{MP21}.\\

{\bf Corollary 2.1.} {\em A binary linear sum-rank-metric code with block length $4^m-1$ and matrix size $2 \times 2$ can be constructed explicitly from two quaternary BCH codes ${\bf C}_1$ and ${\bf C}_2$ of the minimum distances $d_1$ and $d_2$ satisfying $d_1=2d_2$. The minimum sum-rank distance of this sum-rank-metric code is at least $2d_2=d_1$. The dimension of this sum-rank-metric code is at least $$2(2\cdot 4^m-2-m\left(\frac{3d_1}{2}-2\right)).$$}\\

{\bf Proof.} It is well-known that the BCH code of the minimum distance $d_1$ has its dimension at least $4^m-1-m(d_1-1)$, see \cite[Theorem 5.1.7]{HP}. The conclusion follows directly.\\

However the dimension in Corollary 2.1 is only a lower bound. Sometimes the explicit dimension calculation from cyclotomic cosets is better.\\

{\bf Example 2.3.} For $m=2$, $d_2=3$, a binary linear sum-rank-metric code with block length $15$, matrix size $2 \times 2$ and the minimum sum-rank distance at least $6$ is constructed from $[15, 8, 6]_4$ and $[15, 12, 3]_4$ BCH codes. The dimension of this code is $2 \cdot 20$. From Corollary 2.1, the lower bound of the dimension of this sum-rank-metric code is at least $2 \times 16$, since the lower bound on dimensions of BCH codes in Corollary 2.1 is not explicit. This linear sum-rank-metric code is larger than the constructed sum-rank-metric code of the same minimum sum-rank distance $6$, the dimension $2 \cdot 16$ in \cite[Table IV]{MP21}.\\

For $d_2=7$, we have two $[15, 7, 7]_4$ and $[15,1,14]_4$ BCH codes over ${\bf F}_4$. We get a  binary linear sum-rank-metric code with block length $15$, matrix size $2 \times 2$ and the minimum sum-rank distance at least $14$.  The dimension of this code is $2 \cdot 8$. Our code is much larger than the sum-rank BCH code of the dimension $2 \cdot 2$ and the minimum sum-rank distance $14$, constructed in \cite[Table IV]{MP21}.\\

In the following table, we list all dimensions of binary linear sum-rank-metric codes with the block length $15$ constructed from two quaternary BCH $[n,k_i,d_i]$ codes, $i=1,2,$ satisfying $d_2=\lceil \frac{d_1}{2} \rceil$, where $\lceil \cdot\rceil$ denotes the smallest integer greater than $\cdot$. The Singleton-like bounds are also listed. Some of our codes from two quaternary BCH codes are close to the Singleton-like bound. Many of them are larger than sum-rank-metric codes of the same minimum sum-rank distances constructed in \cite{MP21}. In Tables of \cite{MP21}, the dimensions of sum-rank-metric codes were only calculated for some minimum sum-rank distances. We calculated the dimensions of sum-rank-metric codes from Theorem 9 in \cite{MP21} for other minimum sum-rank distances. However it should be indicated that the true dimensions and the true minimum sum-rank distances of sum-rank BCH codes were not calculated explicitly in \cite{MP21}. The entries from \cite{MP21} are lower bounds of dimensions and distances. More BCH-type sum-rank-metric codes were given in the webpage \cite{Lao}.\\

\begin{longtable}{|c|c|c|c|}
	\caption{\label{tab:2-1} \small{The dimensions of binary linear sum-rank-metric codes with the block length $15$ and matrix size $2\times 2$($d_2=\lceil \frac{d_1}{2} \rceil$).}}\\ \hline
	$d_{sr}\geq$&dimension &Table IV, \cite{MP21}&Singleton-like \\ \hline
	$4$& $2 \cdot 24$  & $ 2 \cdot 20$ & $2 \cdot 27$\\ \hline
	$5$ & $2 \cdot 21$  &$ 2 \cdot 18$ & $2 \cdot 26$ \\ \hline
	$6$ &$2 \cdot 20$  &$  2 \cdot 16$ & $2 \cdot 25$ \\ \hline
	$7$ &$2 \cdot 17$ & $2 \cdot 14$ & $2 \cdot 24$ \\ \hline
	$8$ &$ 2\cdot 15$  & $2 \cdot 10$ & $2 \cdot 23$ \\ \hline
	$9$ &$ 2 \cdot 13$ & $2 \cdot 8$  & $2 \cdot 22$ \\ \hline
	$10$ &$2 \cdot 13$ & $2 \cdot 8$  & $2 \cdot 21$ \\ \hline
	$11$ &$2 \cdot 11$ &  $2\cdot 6$  & $2 \cdot 20$ \\ \hline
	$12$ &$2 \cdot 10$ & $2 \cdot 4$  & $2 \cdot 19$ \\ \hline
	$13$ &$ 2 \cdot 8$ &   $2\cdot 2$ & $2\cdot 18$ \\ \hline
	$14$ &$2 \cdot 8$ &  $2\cdot 2$ & $2 \cdot 17$ \\ \hline
	$15$ &$2 \cdot 6$  &    $2\cdot 2$ & $2 \cdot 16$  \\ \hline
\end{longtable}

{\bf Remark 2.2.} It is well-known that for many small parameters, quaternary BCH codes are not the presently known best codes, see \cite{codetable}. Therefore larger binary linear sum-rank-metric codes were constructed in \cite{Chen}.\\

Some lower bounds on the minimum sum-rank distances of sum-rank BCH codes were improved in a recent paper \cite{ALNWZ}. However our linear sum-rank-metric codes from BCH codes over ${\bf F}_4$ are better than sum-rank BCH codes in \cite{MP21,ALNWZ}, in particular, when the minimum sum-rank distances are larger. We speculate that perhaps there are some unknown deep algebraic structures in these sum-rank BCH codes of \cite{MP21}. For example, are automorphism groups of these sum-rank BCH codes of large minimum sum-rank distances larger?\\

In some cases, the minimum sum-rank distance $d_{sr}$ of $SR({\bf C}_1, {\bf C}_2)$ is much larger than our lower bound $\min\{d_1, 2d_2\}$, as showed in the following example.\\

{\bf Example 2.4.} Let ${\bf C}_1$ be a $[25,1,25]$ narrow-sense BCH code over
${\bf F}_4$ with designed distance $\delta_1=15$. Let ${\bf C}_2$ be a
$[25,2,20]$ BCH code over
${\bf F}_4$ with designed distance $\delta_2=10$ and defining set ${\bf T}_2=C_0\cup C_1\cup C_2\cup C_5$. Let $SR({\bf C}_1, {\bf C}_2)$ be a binary linear sum-rank-metric code with block length $25$ and matrix size $2 \times 2$. From SageMath calculation, the minimum sum-rank distance is actually $d_{sr}=30$.\\

It is well-known BCH codes are asymptotically bad. Then it is impossible to get the asymptotically good sequence of sum-rank-metric codes from BCH codes. Asymptotically good sequence of binary linear sum-rank-metric codes will be constructed from Goppa codes in the next section.\\

\subsection{Goppa-type sum-rank-metric codes}

The class of Goppa codes invented by V. D. Goppa in his 1970-1971  papers \cite{Goppa1,Goppa2}, is one of most interesting classes of error-correcting codes and is the origin of algebraic-geometric codes. It is well-known, that there are sequences of binary Goppa codes meeting the Gilbert-Varshamov bound $$R(\delta) \geq 1-H_2(\delta),$$ $H_2(\delta)$ is the entropy function, see \cite[pp. 308-309]{MScode}, \cite[Chapter 12]{MScode} and \cite[Chapter 9]{Lint}.  There is no the sum-rank version of Goppa codes in the Hamming-metric, as linearized Reed-Solomon codes and sum-rank BCH codes in \cite{MP1,MP21}. Then it is interesting to calculate parameters of binary linear sum-rank-metric codes with the matrix size $2 \times 2$ in Theorem 2.1 from binary or quaternary Goppa codes.\\

Let $G(z)$ be a polynomial in ${\bf F}_{q^m}[z]$ and $L=\{\alpha_1, \ldots, \alpha_n\}\subset {\bf F}_{q^m}$ be a subset of ${\bf F}_{q^m}$ satisfying $g(\alpha_i)\neq 0$ for $i=1, \ldots, n$. The Goppa code $\Gamma(L,G)$ is the set of codewords $(c_1, \ldots, c_n) \in {\bf F}_q^n$ satisfying $\Sigma_{i=1}^n\frac{c_i}{z-\alpha_i} \equiv 0$ $mod$ $G(z)$, see \cite{Goppa1,Goppa2}, \cite[Chapter 12]{MScode} and \cite[Chapter 9]{Lint}. This is a subfield subcode of the GRS code with the dimension $$k(\Gamma(L,G))) \geq n-m\deg(G),$$ and the minimum distance $$d(\Gamma(L,G)) \geq \deg(G)+1.$$ Therefore the length is flexible.\\

When $q=2$, this is the binary Goppa code. A binary Goppa code is more interesting since a better lower bound on the minimum distance can be obtained. If the Goppa polynomial $G(z) \in {\bf F}_{2^m}[z]$ has no multiple roots, this is called a separable Goppa code. The minimum distance of a separable binary Goppa code satisfies $$d(\Gamma(L,G)) \geq 2\deg(G)+1,$$ see \cite[Theorem 6, Section3]{MScode}.\\

The typical example of binary separable Goppa codes is the irreducible Goppa code with an irreducible Goppa polynomial $G(z) \in {\bf F}_{2^m}[z]$ of the degree $r$. Then a family of binary Goppa $[2^m, k\geq 2^m-rm, d \geq 2r+1]_2$ codes $\Gamma_{r,m}(L,G)$ was constructed, see page 345 of \cite{MScode}.\\

{\bf Theorem 2.3.} {\em Sequences of binary linear sum-rank-metric codes with matrix size $2 \times 2$ of the rate $$R_{sr}(\delta_{sr}) \geq 1-\frac{1}{2}\left(H_4(\delta_{sr})+H_4(2\delta_{sr})\right),$$ can be constructed from binary Goppa codes, for any $\delta_{sr}$ satisfying $0<\delta_{sr} <\frac{1}{4}$.}\\

{\bf Proof.} We take two sequences of quaternary $[\ell_i, k_i, d_i]_2$ and $[\ell_i, k_i',d_i']_2$ Goppa codes meeting the Gilbert-Varshamov bound, satisfying $2d_i=d_i'$, see \cite[Theorem 9.4.1]{Lint}. Then $$\lim \frac{k_i}{\ell_i} \geq 1-H_4\left(\lim \frac{d_i}{\ell_i}\right),$$ and $$\lim \frac{k_i'}{\ell_i} \geq 1-H_4\left(\lim \frac{d_i'}{\ell_i}\right).$$ From Theorem 2.1, we get a sequence of linear sum-rank-metric codes with the matrix size $2 \times 2$, the sum-rank rate $R_{sr}=\lim \frac{2(k_i+k_i')}{4\ell_i}=\frac{1}{2}\left(\lim \frac{k_i}{\ell_i}+\lim\frac{k_i'}{\ell_i}\right)$. The conclusion follows immediately.\\

In the following table we list some binary linear sum-rank-metric codes with the block length $32$, $64$ and $128$, the matrix size $2 \times 2$, from applying two binary irreducible Goppa codes in Theorem 2.1. These codes constructed from Theorem 2.1 applying to Hamming-metric Goppa codes are called Goppa-type sum-rank-metric codes. Our codes can be comparable with corresponding sum-rank BCH codes in \cite{MP21} of the block size $31$, $63$ and $127$.\\

\begin{longtable}{|l|l|l|l|l|}
\caption{\label{tab:A-q-5-3} The dimensions of binary linear sum-rank-metric codes with the block length $\ell$ and matrix size $2\times 2$.}\\ \hline
$\ell$&$d_{sr}\geq$&dimension &Tables, \cite{MP21}&Singleton-like \\ \hline
$32$&$5$ &$2 \cdot 49$ &$\ell=31$, $\dim=2 \cdot 47$ & $2 \cdot 60$ \\ \hline
$32$&$18$ &$2 \cdot 12$ &$\ell=31$, $\dim=2 \cdot 7$ & $2 \cdot 47$ \\ \hline
$32$&$22$ &$2 \cdot 7$ &$\ell=31$, $\dim=2 \cdot 7$ & $2 \cdot 43$ \\ \hline
$32$&$26$ &$2 \cdot 2$ &$\ell=31$, $\dim=2 \cdot 2$ & $2 \cdot 39$ \\ \hline
$64$&$5$ & $2 \cdot 110$  &$\ell=63$, $\dim=2 \cdot 108$ & $2 \cdot 124$ \\ \hline
$128$&$5$ & $2 \cdot 235$&$\ell=127$, $\dim=2 \cdot 233$ &$2 \cdot 252$\\ \hline
\end{longtable}

It is natural to use two quaternary Goppa codes in Theorem 2.1. Some of them have worse lower bounds on dimensions, compared with corresponding sum-rank BCH codes in \cite{MP21} of the same minimum sum-rank distances. Goppa-type sum-rank-metric codes in the following example can be comparable with sum-rank BCH codes in \cite{MP21}.\\

{\bf Example 2.5.} Let ${\bf C}_2$ be the $[64, 37, 10]_4$ Goppa code over ${\bf F}_4$, another code ${\bf C}_1$ be the $[64, 52, 5]_4$ Goppa code. From Theorem 2.1, a binary linear sum-rank-metric code with the block length $64$, the matrix size $2 \times 2$, the minimum sum-rank distance at least $10$ and the dimension $2 \cdot (37+52)=2 \cdot 89$ is constructed. This code can be comparable with the sum-rank BCH code with the block length $63$, the dimension $2 \cdot 88$ and the same minimum sum-rank distance $10$ in \cite[Table VI]{MP21}. Similarly, a binary linear sum-rank-metric code with the block length $64$, the matrix size $2 \times 2$, the minimum sum-rank distance at least $14$ and the dimension $2 \cdot (25+46)=2 \cdot 71$ is constructed. This code can be comparable with the sum-rank BCH code with the block length $63$, dimension $2 \cdot 70$ and the same minimum sum-rank distance $14$ in \cite[Table VI]{MP21}.\\

\section{Linear sum-rank-metric codes from additive quaternary codes}

Let us recall the following definition of additive quaternary codes in \cite{Blokhuis,Bier,Bier1}, which are the natural extension of quaternary linear codes.\\

{\bf Definition 3.1.} {\em Let ${\bf C} \subset {\bf F}_4^n$ be a general code. Suppose that ${\bf C}$ satisfies the additive closure property, ${\bf c}_1+{\bf c}_2 \subset {\bf C}$, if ${\bf c}_i \in {\bf C}$ for $i=1, 2$. We call this code ${\bf C}$ an additive  quaternary code.}\\

A little more generalization of Theorem 2.1 is the following result. It is obvious that we only need the ``additivity" in the two Hamming-metric codes ${\bf C}_1$ and ${\bf C}_2$ to preserve the ``additivity" of the sum-rank-metric code $SR({\bf C}_1, {\bf C}_2)$.\\

{\bf Theorem 3.1.} {\em Let ${\bf C}_1 \subset {\bf F}_4^\ell$ and ${\bf C}_2 \subset {\bf F}_4^\ell$ be two additive $(t, 4^{k_1}, d_1)_4$ and $(t, 4^{k_2}, d_2)_4$ codes. Then a binary linear sum-rank-metric code $SR({\bf C}_1, {\bf C}_2)$ with block length $\ell$, matrix size $2 \times 2$ and the minimum sum-rank distance at least $\min\{d_1, 2d_2\}$ can be constructed explicitly. The dimension of $SR({\bf C}_1, {\bf C}_2)$ over ${\bf F}_2$ is $\dim_{{\bf F}_2}(SR({\bf C}_1, {\bf C}_2))=2(k_1+k_2)$.}\\

It should be noticed that the ``dimension" $k$ of an additive quaternary code may be a positive integer or the half of a positive integer, see \cite{Bier}. Comparing with the presently known best quaternary linear codes in \cite{codetable}, some larger additive quaternary codes were constructed in \cite{Blokhuis,Bier,Bier1,Bier2}. Then the binary linear sum-rank-metric codes constructed in \cite{Chen} can be improved as the following examples.\\

{\bf Example 3.1.} We want to construct a binary linear sum-rank-metric code with the block length $12$, the matrix size $2 \times 2$ and the minimum sum-rank distance $8$. If we use Theorem 2.1 or 2.2, two linear $[12, k_1, 4]_4$ and $[12, k_2, 8]_4$ codes are needed. From \cite{codetable}, the largest such linear quaternary codes are $[12, 8, 4]_4$ and $[12, 3, 8]_4$ codes. The dimension of binary linear sum-rank-metric code constructed in Theorem 2.1 or 2.2 is of the dimension $2 \cdot (3+8)=22$. However from Table 1 of \cite{Bier1}, the $(12, 4^{3.5}, 8)_4$ additive quaternary code was conctructed. We can construct a binary linear sum-rank-metric code of the dimension $2\cdot (3.5+8)=23$, with the same minimum sum-rank distance $8$.\\

Similarly, if we want to construct a binary linear sum-rank-metric code with the block length $13$, the matrix size $2 \times 2$ and the minimum sum-rank distance $10$ from Theorem 2.1 or 2.2, and the presently known largest codes in \cite{codetable}, then the dimension is $2\cdot (7+2)=18$. However from Table 1 of \cite{Bier1}, the  $(13, 4^{7.5}, 5)_4$ additive quaternary code was constructed. We can construct a binary linear sum-rank-metric code with the dimension $2\cdot (7.5+2)=19$ and the same minimum sum-rank distance $10$.\\

{\bf Example 3.2.} Additive $(19, 4^{14.5},4)_4$, $(20, 4^{15.5},4)_4$,$(21, 4^{16.5},4)_4$ and $(22, 4^{17.5},4)_4$ codes were constructed in \cite{Bier2}. However the largest dimensions of linear quaternary codes of length $19+i$, $i=0,1,2,3$, and the minimum distance $4$, are $14, 15, 16, 17$, see \cite{codetable}. The largest dimensions of linear quaternary codes of length $19, 20, 21, 22$, and the minimum distance $8$, are $9, 10, 10, 11$, see \cite{codetable}. Therefore the binary linear sum-rank-metric codes with the matrix size $2 \times 2$ and the block length $19, 20, 21, 22$ constructed in Theorem 2.1 or 2.2 have their dimensions $2 \cdot(9+14)=46$, $2 \cdot(15+10)=50$, $2 \cdot (16+10)=52$ and $2 \cdot (17+11)=56$.\\

From Theorem 3.1 and additive quaternary codes in \cite{Bier2}, we construct binary linear sum-rank-metric codes with the block length $19, 20, 21, 22$, the matrix size $2 \times 2$, the minimum sum-rank distance $10$ and the dimensions $47,51,53,57$.\\

More BCH-type binary linear sum-rank-metric codes and binary linear sum-rank-metric codes from additive quaternary codes are listed in the webpage \cite{Lao} for the convenience of readers.\\

\section{Fast decoding of binary linear sum-rank-metric codes with the matrix size $2 \times 2$}

In this section, we first give a reduction of the decoding in the binary sum-rank-metric space to several decodings in the Hamming-metric space. Then our reduction method is applied to BCH-type and Goppa-type binary linear sum-rank-metric codes with the matrix size $2 \times 2$.\\

\subsection{Reduction of the decoding in the binary sum-rank-metric space to the decoding in the Hamming-metric space}

The key point of the reduction of the decoding in the binary sum-rank-metric space to the decoding in the Hamming-metric space is the identification $${\bf F}_q^{n\times n}\backsimeq\{a_0x+a_1x^q+\cdots+a_{n-1}x^{q^{n-1}}: a_0, \ldots, a_{n-1} \in {\bf F}_{q^n}\}.$$ This is the isomorphism of the linear space over ${\bf F}_q$. In the case $q=2, m=n=2$, it is of the following form, $${\bf F}_2^{2\times 2}\backsimeq\{a_0x+a_1x^2: a_0, a_1 \in {\bf F}_4 \}.$$  In this paper our decoding algorithms only work for these binary linear sum-rank-metric codes of the matrix size $2 \times 2$ from one quaternary code ${\bf C}_2$ and one quaternary code ${\bf C}_1$. We give our reduction method to reduce the decoding of the constructed binary linear sum-rank-metric codes to the decoding of the Hamming-metric codes ${\bf C}_1$ and ${\bf C}_2$.\\

Let ${\bf v}=(v_1, \ldots, v_\ell) \in {\bf F}_4^\ell$ be a Hamming weight $\ell$ vector, that is, $v_i \neq 0$, $i=1, \ldots, \ell$,  and ${\bf C} \subset {\bf F}_4^\ell$ be a quaternary linear  code, then ${\bf v} \cdot {\bf C} \subset {\bf F}_4^\ell$ is the quaternary linear code defined by $${\bf v} \cdot {\bf C}=\{(v_1c_1, \ldots,v_\ell c_\ell): (c_1, \ldots, c_\ell) \in {\bf C}\}.$$ We always represent ${\bf F}_4=\{0,1,\omega, \omega^2\}$, where $\omega^2+\omega+1=0$. Set ${\bf 1}=(1, \ldots, 1) \in {\bf F}_4^\ell$, ${\bf \omega}=(\omega, \ldots, \omega) \in {\bf F}_4^\ell$, and ${\bf \omega}^2=(\omega^2, \ldots, \omega^2) \in {\bf F}_4^\ell$.\\

For any received vector ${\bf y} \in {\bf F}_2^{2\times 2, \ldots, 2\times 2}$, of the form $${\bf y}={\bf c}+{\bf e},$$ where ${\bf c} \in SR({\bf C}_1, {\bf C}_2)$, and $$wt_{sr}({\bf e}) \leq \left \lfloor \frac{d_{sr}-1}{4} \right \rfloor,$$£¬ it is clear that the decoding of the sum-rank-metric codes constructed in Theorem 2.1 from two Hamming-metric error-correcting codes ${\bf C}_1$ and ${\bf C}_2$ satisfying $d_1 \geq d_{sr}$ and $d_2 \geq \frac{d_{sr}}{2}$, can be reduced to the decoding of these two codes ${\bf C}_1$ and ${\bf C}_2$. However if we want to decode sum-rank-metric codes up to $$wt_{sr}({\bf e}) \leq \left \lfloor \frac{d_{sr}-1}{2} \right \rfloor,$$ errors, our reduction method requires  that the two quaternary linear $[\ell, k_1, d_1]_4$ code ${\bf C}_1$ and the $[\ell, k_2, d_2]_4$ code ${\bf C}_2$  satisfy the condition, $d_1=d_{sr}$ and $d_2 \geq \frac{2d_{sr}}{3}$. \\

For any received vector ${\bf y} \in {\bf F}_2^{2\times 2, \ldots, 2\times 2}$, of the form $${\bf y}={\bf c}+{\bf e},$$ where ${\bf c} \in SR({\bf C}_1, {\bf C}_2)$, and $$wt_{sr}({\bf e}) \leq \left \lfloor \frac{d_{sr}-1}{2} \right \rfloor,$$ we represent each coordinate matrix of ${\bf y}$ at the $i$-th position $y_i=y_{0,i}x+y_{1,i}x^2=c_{2,i}x+c_{1,i}x^2+e_{0,i}x+e_{1,i}x^2 \in {\bf F}_2^{2\times 2}$. That is $${\bf y}={\bf y}_0x+{\bf y}_1x^2=({\bf c}_2+{\bf e}_0)x+({\bf c}_1+{\bf e}_1)x^2,$$ where ${\bf c}_1=(c_{1,1}, \ldots, c_{1, \ell}) \in {\bf C}_1$, ${\bf c}_2=(c_{2,1}, \ldots, c_{2,\ell}) \in {\bf C}_2$, ${\bf e}_0, {\bf e}_1 \in {\bf F}_4^\ell$. Let us observe the sum-rank weight of the error vector ${\bf e}={\bf e}_0x+{\bf e}_1x^2$, then $$wt_{sr}({\bf e})=2i_1+2i_2+i_3,$$ where $i_1$ is the number of the coordinate positions in the subset $I_1 \subset \{1, \ldots, \ell\}$ such that $e_0 \neq 0$ and $e_1 =0$, $i_2$ is the number of the coordinate positions in the subset $I_2 \subset \{1, \ldots, \ell\}$ such that $e_0=0$ and $e_1 \neq 0$, and $i_3$ is the number of the coordinate positions in the subset $I_3 \subset \{1, \ldots, \ell\}$ such that $e_0 \neq 0$ and $e_1 \neq 0$. It is obvious that the Hamming weights of ${\bf e}_0$ and ${\bf e}_1$ satisfy $wt_H({\bf e}_0)=i_1+i_3 \leq wt_{sr}({\bf e})$ and $wt_H({\bf e}_1)=i_2+i_3 \leq wt_{sr}({\bf e})$.\\

First of all we decode the received word ${\bf y}_1 \in {\bf F}_4^\ell$. Since ${\bf y}_1={\bf c}_1+{\bf e}_1$ and $$wt_H({\bf e}_1) \leq \left \lfloor \frac{d_{sr}-1}{2} \right \rfloor=\left \lfloor \frac{d_1-1}{2} \right \rfloor,$$ then any decoding algorithm correcting up to half distance of ${\bf C}_1$ can decode ${\bf y}_1$ to the form ${\bf y}_1={\bf c}_1+{\bf e}_1$. Therefore ${\bf e}_1$ is determined from the decoding algorithm of ${\bf C}_1$.\\

We get another ``received word" ${\bf y}'={\bf y}+{\bf c}_1x^2=({\bf c}_2+{\bf e}_0)x+{\bf e}_2x^2$. Though we do not know what the nonzero element $\frac{e_{0,i}}{e_{1,i}}$ in ${\bf F}_4$ is for the coordinate position $i \in I_3$. There are three possibilities $1, \omega, \omega^2$ for this nonzero element for each position $i \in I_3$. By applying the ``received word" ${\bf y}'$ to three vectors ${\bf 1}$, ${\bf \omega}$ and ${\bf \omega}^2$, we get $${\bf c}_2+({\bf e}_0+{\bf e}_1),$$  $${\bf \omega} \cdot {\bf c}_2+({\bf e}_0 \cdot {\bf \omega}+{\bf e}_1 \cdot {\bf \omega}^2),$$ and $${\bf \omega}^2 \cdot {\bf c}_2+({\bf e}_0 \cdot {\bf \omega}^2+{\bf e}_1 \cdot {\bf \omega}).$$ Therefore for $i \in I_3$, $$e_{0,i}x+e_{1,i}x^2$$ is zero for at least $\frac{i_3}{3}$ coordinate positions, when $x$ takes one of three possibilities $x=1, \omega, \omega^2$. Then the Hamming weight of one of three vectors $$({\bf e}_0+{\bf e}_1),$$  $$({\bf e}_0 \cdot {\bf \omega}+{\bf e}_1 \cdot {\bf \omega}^2),$$ and $$({\bf e}_0 \cdot {\bf \omega}^2+{\bf e}_1 \cdot {\bf \omega}).$$ is at most $$i_1+i_2+\frac{2i_3}{3}=\frac{wt_{sr}({\bf e})}{2}+\frac{i_3}{6} \leq \frac{2wt_{sr}({\bf e})}{3} \leq \left \lfloor \frac{d_2-1}{2} \right \rfloor,$$ from the assumption $d_2 \geq \frac{2d_{sr}}{3}$.\\

By applying any decoding algorithm of the code ${\bf C}_2$, correcting up to $\left \lfloor \frac{d_2-1}{2} \right \rfloor$ errors, to three equivalent codes ${\bf C}_2$, ${\bf \omega} \cdot {\bf C}_2$ and ${\bf \omega}^2 \cdot {\bf C}_2$, on three ``received words" $${\bf y}'(1)={\bf y}({\bf 1})+{\bf c}_1,$$ $${\bf y}'({\bf \omega})={\bf y}({\bf \omega})+{\bf \omega}^2 \cdot {\bf c}_1,$$ and $${\bf y}'({\bf \omega}^2)={\bf y}({\bf \omega}^2)+{\bf \omega} \cdot {\bf c}_1,$$ we can decode to determine the whole error ${\bf e}$.\\

{\bf Theorem 4.1.} {\em Let $SR({\bf C}_1, {\bf C}_2)$ be a binary linear sum-rank-metric code with block length $\ell$, matrix size $2 \times 2$ and the minimum sum-rank distance $d_{sr}$ constructed from two quaternary $[\ell, k_i, d_i]_4$ codes ${\bf C}_i$, $i=1,2$ satisfying $d_1 \geq d_{sr}$ and $d_2 \geq \frac{2d_{sr}}{3}$ as in Theorem 2.1. Suppose that there are decoding algorithms correcting up to $\left \lfloor \frac{d_i-1}{2} \right \rfloor$ errors for the code ${\bf C}_i$ with the cost ${\bf T}_i$, $i=1, 2$. Then by applying the decoding algorithm to the code ${\bf C}_1$ once,  and the the decoding algorithm to three equivalent codes ${\bf C}_2$, ${\bf \omega} \cdot {\bf C}_2$ and ${\bf \omega}^2 \cdot {\bf C}_2$ three times, we can decode the sum-rank-metric code $SR({\bf C}_1, {\bf C}_2)$ up to $\left \lfloor \frac{d_{sr}-1}{2} \right \rfloor$ errors. The cost of this decoding algorithm in the sum-rank-metric is $${\bf T}_1+3{\bf T}_2.$$}\\

It is obvious that binary linear sum-rank-metric codes constructed in Theorem 2.2 can be decoded similarly, if the condition $d_1 \geq \frac{2d_{sr}}{3}$ is assumed. The following examples show that our reduction of the decoding in the sum-rank-metric to the decoding in the Hamming-metric works well. For the convenience of the readers, writing ${\bf x}_a$ for a vector part that consists of $a$ $x$'s for each $x\in {\bf F}_4$.\\

{\bf Example 4.1.}  Let ${\bf C}_1$ be a $[63,22,24]$ BCH code over
${\bf F}_4$ with designed distance $\delta_1=24$ and defining set ${\bf T}_1=C_0\cup C_1\cup C_2\cup C_3\cup C_5\cup C_6\cup C_7\cup C_9\cup C_{10}\cup C_{11}\cup C_{13}\cup C_{14}\cup C_{15}\cup C_{21}\cup C_{22}$. Let ${\bf C}_2$ be a
$[63,29,16]$ BCH code over
${\bf F}_4$ with designed distance $\delta_2=16$ and defining set ${\bf T}_2=C_0\cup C_1\cup C_2\cup C_3\cup C_5\cup C_6\cup C_7\cup C_9\cup C_{10}\cup C_{11}\cup C_{13}\cup C_{14}$.  Let $SR({\bf C}_1, {\bf C}_2)$ be a binary linear sum-rank-metric code with block length $63$, matrix size $2 \times 2$ and the minimum sum-rank distance $d_{sr}\ge24$. Take
$\mathbf{c}_1=(0, \omega, \omega + 1, \omega, \{\bm{\omega + 1}\}_2, \omega, 1, 0, \bm{\omega}_2, 0, \omega, 0, \bm{\omega}_3, {\bf 0}_2, \omega, 0, 1, \omega + 1, \omega, {\bf 1}_2, 0, \omega + 1, \omega, 1, \omega, \omega + 1, 0, \omega + 1, \bm{\omega}_2, \omega + 1, {\bf 1}_2, \omega, 1, \omega, 1, 0, \omega, 1, \omega + 1, 0, \omega, 1, \omega, 1, \omega + 1, \omega, 0, \omega + 1, 1, 0, \omega, {\bf 1}_2, \omega, 0)\in \mathbf{C}_1$,
$\mathbf{c}_2=({\bf 1}_2, {\bf 0}_2, \omega + 1, 0, 1, \omega, \omega + 1, \bm{\omega}_2, \omega + 1, \omega, \omega + 1, 0, 1, \bm{\omega}_2, 1, \bm{\omega}_2, 1, 0, \bm{\omega}_2, {\bf 0}_2, \omega + 1, 1, 0, \omega, {\bf 1}_2, \omega + 1, \omega, \{\bm{\omega + 1}\}_2, 0, 1, \omega, \{\bm{\omega + 1}\}_2, \omega, \omega + 1, 0, \omega + 1, \omega, 1, \bm{\omega}_2, 0, 1, \omega + 1, {\bf 0}_2, \omega + 1, 1, 0, 1, 0, \{\bm{\omega + 1}\}_2, \omega)\in \mathbf{C}_2$, $\mathbf{e}_1=(1, 0, \omega + 1, 0, \omega, 0, 1, 0, \omega + 1, 0, \omega, 0, 1, 0, \omega + 1, 0, \omega, 0, 1, 0, \omega + 1, {\bf 0}_{42})$, and
$\mathbf{e}_2=(1, 0, 1, 0, 1, 0, 1, 0, 1, 0, 1, 0, 1, 0, 1, 0, 1, 0, 1, 0, 1, {\bf 0}_{42})$. Then $wt_{H}(\mathbf{e}_1)=11$, $wt_{H}(\mathbf{e}_2)=11$, and
$wt_{st}(\mathbf{e})=11\leq \lfloor{\frac{d_{sr}-1}{2}}\rfloor$. Using the decoding command in SageMath, we can get $\mathbf{c}_1$ from
$\mathbf{y}=\mathbf{c}_1+\mathbf{e}_1$. Since $wt_H(\mathbf{e}_2)=11>\lfloor \frac{16-1}{2} \rfloor=7$, we can not get $\mathbf{c}_2$ from the decoding $\mathbf{c}_2+\mathbf{e}_2$. We can decode three words
$\mathbf{y}'(1)$, $\mathbf{y}'(\omega)$ and
$\mathbf{y}'(\omega^2)$, and get $ \mathbf{c}_2$ from
$\mathbf{y}'(1)$  with the Hamming weight $wt_{H}(\mathbf{e}_2+ \mathbf{e}_1)=7$ of the error vector. Hence, we decode the above received word of $SR({\bf C}_1, {\bf C}_2)$. \\
%

\subsection{Fast decoding algorithms of BCH-type binary linear sum-rank-metric codes}

The following dimensions of binary linear sum-rank-metric codes with the block length $15$ and the matrix size $2 \times 2$ are constructed from two quaternary BCH $[\ell, k_i, d_i]_4$ codes, $i=1,2$ satisfying $d_1=d_{sr}$ and $d_2 = \lceil \frac{2d_{sr}}{3} \rceil$. From Theorem 2.1, the constructed binary linear sum-rank-metric codes have their minimum sum-rank distances at least $\min\{d_1, 2d_2\}=d_{sr}$. Though these codes are smaller than BCH-type sum-rank-metric codes in Section 2. Comparing to these sum-rank BCH codes in \cite{MP21}, our codes are larger when the minimum sum-rank distances are the same. Therefore it is good to give fast decoding algorithms of  these BCH-type binary linear sum-rank-metric codes.\\

\begin{longtable}{|c|c|c|c|}
\caption{\label{tab:4-1} \small{The dimensions of binary linear sum-rank-metric codes with the block length $15$ and matrix size $2\times 2$($d_2 = \lceil \frac{2d_{sr}}{3} \rceil$).}}	\\ \hline
	$d_{sr}\geq$&dimension &Table IV, \cite{MP21}&Singleton-like \\ \hline
	$4$& $2 \cdot 22$  & $2 \cdot 20$ & $2 \cdot 27$\\ \hline
	$5$ & $2 \cdot 19$  & $2 \cdot 18$ & $2 \cdot 26$ \\ \hline
	$6$ &$2 \cdot 18$  & $2 \cdot 16$ & $2 \cdot 25$ \\ \hline
	$7$ &$2 \cdot 16$ & $2 \cdot 14$ & $2 \cdot 24$ \\ \hline
	$8$ &$ 2\cdot 13$  & $2 \cdot 10$ & $2 \cdot 23$ \\ \hline
	$9$ &$ 2 \cdot 12$ & $2 \cdot 8$  & $2 \cdot 22$ \\ \hline
	$10$ &$2 \cdot 11$ & $2 \cdot 8$  & $2 \cdot 21$ \\ \hline
	$11$ &$2 \cdot 8$ & $2\cdot 6$  & $2 \cdot 20$ \\ \hline
	$12$ &$2 \cdot 7$ & $2 \cdot 4$  & $2 \cdot 19$ \\ \hline
	$13$ &$ 2 \cdot 5$ & $2 \cdot 2$  & $2\cdot 18$ \\ \hline
	$14$ &$2 \cdot 5$ & $2\cdot 2$ & $2 \cdot 17$ \\ \hline
	$15$ &$2 \cdot 5$  &   $2 \cdot 2$ & $2 \cdot 16$  \\ \hline
\end{longtable}

In general, the binary linear sum-rank-metric codes of the block length $4^h-1$, the matrix size $2 \times 2$, constructed from two BCH quaternary codes ${\bf C}_1$ and ${\bf C}_2$ satisfying $d_1 \geq d_{sr}$ and $d_2 \geq \frac{d_{sr}}{3}$ as in Theorem 2.1, are larger than sum-rank BCH codes in \cite{MP21} of the same minimum sum-rank distances. For these better BCH-type binary linear sum-rank-metric codes, we have the following result from Theorem 4.1.\\

{\bf Corollary 4.1.} {\em Let $SR({\bf C}_1, {\bf C}_2)$ be a BCH-type binary linear sum-rank-metric code with block length $\ell$, matrix size $2 \times 2$ and the minimum sum-rank distance $d_{sr}$ constructed from two quaternary BCH $[\ell, k_i, d_i]_4$ codes ${\bf C}_i$, $i=1,2$ as in Theorem 2.1, satisfying $d_1 \geq d_{sr}$ and $d_2 \geq \frac{2d_{sr}}{3}$. Then by applying any BCH decoding algorithm to the BCH code ${\bf C}_1$ once and the three equivalent BCH codes ${\bf C}_2$, ${\bf \omega} \cdot {\bf C}_2$ and ${\bf \omega}^2 \cdot {\bf C}_2$ three times, we can decode the sum-rank-metric code $SR({\bf C}_1, {\bf C}_2)$ up to $\left \lfloor \frac{d_{sr}-1}{2} \right \rfloor$ errors.}\\

Therefore for all above BCH-type linear sum-rank-metric codes better than sum-rank BCH codes invented in \cite{MP21}, we give the decoding algorithm, which can correct up to $\left \lfloor \frac{d_{sr}-1}{2} \right \rfloor$ errors. The complexity is the same as the Welch-Berlemkamp algorithm given in \cite{MP1,MP21}. Since our decoding in the sum-rank-metric is reduced to the decoding in the Hamming-metric, it seems that our decoding can be implemented practically by the using of fast BCH decoders in the Hamming-metric.\\

\subsection{Fast decoding algorithms of Goppa-type binary linear sum-rank-metric codes}

By applying Theorem 4.1 and the Sugiyama decoding algorithm, see \cite{Sugiyama} and \cite[Section 5.4.3]{HP}, to the binary linear sum-rank-metric codes with the matrix size $2 \times 2$, constructed from two Goppa binary or quaternary codes satisfying $d_1=d_{sr}$ and $d_2 \geq \frac{2d_{sr}}{3}$ as in Theorem 2.1, we get the fast decoding algorithm of Goppa-type binary linear sum-rank-metric codes with the matrix size $2 \times 2$. These Goppa-type binary linear sum-rank-metric codes can be decoded up to $\left \lfloor \frac{d_{sr}-1}{2} \right \rfloor$ errors, within the complexity $O(\ell^2)$ operations in ${\bf F}_4$. \\

On the other hand, a little weaker version of asymptotically good binary linear sum-rank-metric codes with the matrix size $2 \times 2$ in Theorem 2.3 can be fastly decoded to $\left \lfloor \frac{d_{sr}-1}{2} \right \rfloor$ errors, within the complexity $O(\ell^2)$ operations in ${\bf F}_4$. \\

{\bf Theorem 4.2.} {\em Sequences of binary linear sum-rank-metric codes with the matrix size $2 \times 2$  and the rate $$R_{sr}(\delta_{sr}) \geq 1-\frac{1}{2}(H_4\left(\frac{4}{3}\delta_{sr}\right)+H_4(2\delta_{sr})),$$ can be constructed from binary Goppa codes, for any $\delta_{sr}$ satisfying $0<\delta_{sr} <\frac{1}{4}$. These binary linear sum-rank-metric codes can be decoded up to $\left \lfloor \frac{d_{sr}-1}{2} \right \rfloor$ errors, within the complexity $O(\ell_i^2)$ operations in ${\bf F}_4$, where $\ell_i$ is the block length of the sum-rank-metric code.}\\

\subsection{Fast decoders in the binary sum-rank-metric space from fast decoders in the Hamming-metric space}

If we use two length $31$ quaternary BCH $[31, k_i, d_i]_4$ codes, $i=1, 2$,  satisfying $d_1 \geq d_{sr}$ and $d_2 \geq \frac{2d_{sr}}{3}$ in Theorem 2.1 to construct binary linear sum-rank-metric codes of the block length $31$, the dimension is not bigger than that of the corresponding sum-rank BCH codes. For example, when $d_{sr}=12$, we get a dimension $2 \times 25$ binary linear sum-rank-metric code from the $[31, 15, 8]_4$ BCH code and the $[31, 10, 12]_4$ BCH code. The sum-rank BCH code of the same minimum sum-rank distance $d_{sr}=12$ in Table V of \cite{MP21} has the same dimension $2 \times 25$. From Theorem 4.1, our fast decoder for our code has the same complexity as the Welch-Berlemkamp decoder in \cite{MP21}. However by combining various fast decoders in the Hamming-metric, actually we can decode larger binary linear sum-rank-metric code of the same minimum sum-rank distance with the same complexity.\\

By padding zero to the extended QR (quadratic residue) quaternary $[30, 15, 12]_4$ code, we get a quaternary $[31, 15, 12]_4$ code. Then by applying Theorem 2.1 to this code and the quaternary BCH $[31, 15, 8]_4$ code, a binary linear sum-rank-metric code with the block length $31$, the matrix size $2 \times 2$ and the minimum sum-rank distance $12$ is constructed. The dimension of this code over ${\bf F}_2$ is $2 \cdot (15+15)=2 \cdot 30$. This code is larger than the corresponding dimension $2 \cdot25$ sum-rank BCH code of the same minimum sum-rank distance $12$ in Table V of \cite{MP21}.  From Theorem 4.1, by combining the Welch-Berlemkamp decoder for the extended QR code and the quaternary BCH code, the fast decoder for above our constructed binary linear sum-rank-metric code of the dimension $2 \cdot 30$ can be obtained. This decoding algorithm has the same complexity as the Welch-Berlekamp decoder in \cite{MP21}. In \cite{Chen}, we showed that binary linear sum-rank-metric codes of the matrix size $2 \times 2$ from presently best known codes in \cite{codetable} are much larger than sum-rank BCH codes of the same minimum sum-rank distances. If the second Hamming-metric quaternary codes in Tables 1-10 of \cite{Chen} are replaced by a quaternary code satisfying $d_2 \geq \frac{2d_{sr}}{3}$, larger binary linear sum-rank-metric codes than sum-rank BCH codes with the same minimum sum-rank distances can still be constructed explicitly. By applying our reduction result Theorem 4.1, it leads to fast decoding algorithms for these better binary linear sum-rank-metric codes with the matrix size $2 \times 2$, if fast decoders for these presently known best codes in \cite{codetable} are available. As far as we know, there are fast decoding algorithms for many presently known best Hamming-metric codes in \cite{codetable}.\\

This example illustrates that Theorem 4.1 can be applied to better sum-rank-metric codes constructed from various Hamming-metric codes with fast decoding algorithms. Moreover, considering the the implementation using the FFT (fast Fourier transformation), it seems that our decoding algorithms in the sum-rank-metric is more practical than these decoding algorithms developed in \cite{MP1,MP21,Bartz}. It should be remarked that Theorem 4.1 is only for binary linear sum-rank-metric codes with the matrix size $2 \times 2$.\\

\section{Conclusion}

This paper is concentrated on the construction and fast decoding of binary linear sum-rank-metric codes of the matrix size $2 \times 2$. Better binary linear sum-rank-metric codes with the matrix size $2 \times 2$ from additive quaternary codes were given. We gave a general reduction of the decoding in the binary sum-rank-metric space to the decoding in the Hamming-metric space and showed that the decoding of BCH-type and Goppa-type binary linear sum-rank-metric codes can be reduced to the decoding of quaternary BCH and Goppa codes in the Hamming-metric. Then fast decoding algorithms of these BCH-type and Goppa-type binary sum-rank-metric codes with better parameters have been presented. Asymptotically good sequences of quadratic-time encodable and decodable binary linear sum-rank-metric codes with the matrix size $2 \times 2$ have also been constructed. It is an interesting open problem that if the decoding of the sum-rank-metric codes from Theorem 2.1, can be reduced to the decodings of ${\bf C}_1$ and ${\bf C}_2$, without the assumption $d_2 \geq \frac{2d_{sr}}{3}$.\\

{\bf Acknowledgement.} We thank both reviewers and the AE, Professor Diego Napp, sincerely for their helpful comments and suggestions, which improved the presentation of this paper significantly.\\

\end{document}